\documentclass[journal]{IEEEtran}
\usepackage{graphicx,color}

\usepackage[OT1]{fontenc} 
\usepackage[numbers,sort&compress]{natbib}
\usepackage[cmex10]{amsmath}
\usepackage{amssymb}
\usepackage{bm}
\usepackage{braket}
\usepackage{graphicx}
\usepackage{subfigure}
\usepackage{tabularx}
\usepackage{arydshln}
\usepackage{mathtools}
\usepackage{multirow}
\usepackage{algorithm,algorithmic}
\usepackage{url}
\usepackage{listings}
\usepackage{comment}
\usepackage{hyperref}
\usepackage[absolute,overlay]{textpos}
\def\cV{\mathcal{V}}
\def\cE{\mathcal{E}}
\def\cI{\mathcal{I}}

\makeatletter
\def\ps@IEEEtitlepagestyle{%
  \def\@oddfoot{\mycopyrightnotice}%
  \def\@oddhead{\hbox{}\@IEEEheaderstyle\leftmark\hfil\thepage}\relax
  \def\@evenhead{\@IEEEheaderstyle\thepage\hfil\leftmark\hbox{}}\relax
  \def\@evenfoot{}%
}
\def\mycopyrightnotice{%
  \begin{minipage}{\textwidth}
  \centering \scriptsize
\textcopyright 2024 IEEE. Personal use of this material is permitted.
  Permission from IEEE must be obtained for all other uses, in any current or future
  media, including reprinting/republishing this material for advertising or promotional
  purposes, creating new collective works, for resale or redistribution to servers or
  lists, or reuse of any copyrighted component of this work in other works.
  DOI: \href{https://doi.org/10.1109/TQE.2024.3393437}{10.1109/TQE.2024.3393437}
  \end{minipage}
}
\makeatother
\begin{document}
\title{\huge Accelerating Grover Adaptive Search: Qubit and Gate Count Reduction Strategies with Higher-Order Formulations}

\author{Yuki~Sano,
Kosuke~Mitarai,
Naoki~Yamamoto, and Naoki~Ishikawa,~\IEEEmembership{Senior Member,~IEEE}\thanks{Y.~Sano and N.~Ishikawa are with the Faculty of Engineering, Yokohama National University, Yokohama, 240-8501 Kanagawa, Japan (e-mail: ishikawa-naoki-fr@ynu.ac.jp). K.~Mitarai is with the Graduate School of Engineering Science, Osaka University, 1-3 Machikaneyama, Toyonaka, 560-0043 Osaka, Japan. N.~Yamamoto is with the Department of Applied Physics and Physico-Informatics, Keio University, Hiyoshi 3-14-1, Kohoku, Yokohama, Japan.
The work of Y.~Sano was partially supported by the Exploratory Target Project of the Information-technology Promotion Agency (IPA), Japan. The work of K.~Mitarai was partially supported by MEXT Quantum Leap Flagship Program (Grant No. JPMXS0118067394 and JPMXS0120319794), JST COI-NEXT (Grant No. JPMJPF2014), JST PRESTO (Grant No. JPMJPR2019) and JSPS KAKENHI (Grant No. 23H03819). The work of N.~Yamamoto was partially supported by MEXT Quantum Leap Flagship Program (Grant No. JPMXS0118067285 and JPMXS0120319794). The work of N.~Ishikawa was partially supported by the JSPS KAKENHI (Grant No. 22H01484).}}

%\markboth{\today}
%{Shell \MakeLowercase{\textit{et al.}}: Bare Demo of IEEEtran.cls for Journals}
\maketitle

%ARXIV
\TPshowboxesfalse
\begin{textblock*}{\textwidth}(45pt,10pt)
\footnotesize
\centering
Accepted for publication in IEEE Transactions on Quantum Engineering. This is the author's version which has not been fully edited and content may change prior to final publication. Citation information: DOI 10.1109/TQE.2024.3393437
\end{textblock*}
%ARXIV

\begin{abstract}
Grover adaptive search (GAS) is a quantum exhaustive search algorithm designed to solve binary optimization problems. In this paper, we propose higher-order binary formulations that can simultaneously reduce the numbers of qubits and gates required for GAS. Specifically, we consider two novel strategies: one that reduces the number of gates through polynomial factorization, and the other that halves the order of the objective function, subsequently decreasing circuit runtime and implementation cost. Our analysis demonstrates that the proposed higher-order formulations improve the convergence performance of GAS by both reducing the search space size and the number of quantum gates. Our strategies are also beneficial for general combinatorial optimization problems using one-hot encoding.
\end{abstract}


\IEEEpeerreviewmaketitle

\section{Introduction}
Given the concerns regarding the semiconductor miniaturization limit \cite{waldrop2016chips, theis2017end}, 
quantum computing technology is anticipated to have a significant impact on scientific fields such as cryptography, quantum chemical calculation, and combinatorial optimization \cite{humble2018consumer}.
As for quantum combinatorial optimization, there are two major approaches: the quantum approximate optimization algorithm (QAOA) \cite{farhi2014quantum} and the Grover adaptive search (GAS) \cite{gilliam2021grover}.
QAOA depends on noisy intermediate-scale quantum devices, whereas GAS benefits from the realization of future fault-tolerant quantum computing (FTQC).

In both classical and quantum computing, conventionally, combinatorial optimization problems have been formulated as quadratic unconstrained binary optimization (QUBO) problems \cite{LUCAS2014ISING}.
QUBO problems are supported by well-known high-performance mathematical programming solvers such as the CPLEX Optimizer\footnote{https://www.ibm.com/products/ilog-cplex-optimization-studio/cplex-optimizer} and Gurobi Optimizer\footnote{https://www.gurobi.com/solutions/gurobi-optimizer/}, which use the branch and bound algorithm designed for classical computing.
Additionally, the semidefinite relaxation technique \cite{luo2010semidefinite} can be used to obtain a suboptimal solution in polynomial time.
Quantum annealing \cite{kadowaki1998quantum} and coherent Ising machines \cite{inagaki2016coherent} also support QUBO problems.
Given their attractive demonstrations \cite{ohyama2023quantum,kurasawa2021highspeed}, both could potentially benefit a range of industrial applications.

Unlike the QUBO case, when solving a higher-order unconstrained binary optimization (HUBO) problem in classical computing, it is common to add auxiliary variables and reformulate the problem into a QUBO or an integer programming problem.
Here, the addition of auxiliary variables exponentially enlarges the search space and makes it more difficult to obtain an optimal solution, thus increasing the importance of an efficient solver designed for HUBO problems \cite{valiante2021computational}.
Since the interaction between qubits is not limited to two, in quantum computing, a HUBO formulation arises as a natural approach.
QAOA is an established quantum algorithm that can deal with a HUBO problem \cite{campbell2022qaoa} and is particularly useful when an approximated solution is sufficient.
In contrast, GAS \cite{gilliam2021grover} is an exhaustive search algorithm that has the potential for finding a global optimal solution of a HUBO problem, which is currently the one and only approach in quantum computing.

Assuming FTQC, GAS provides a quadratic speedup for solving a QUBO or HUBO problem \cite{gilliam2021grover}.
Specifically, for $n$ binary variables, $O(2^n)$ evaluations are required in the classical exhaustive search, whereas the query complexity of GAS is $O(\sqrt{2^n})$, and this improvement is termed quadratic speedup.
In conventional studies, the major challenge for Grover-based algorithms is the construction of a quantum circuit for computing an objective function.
To address this challenge, Gilliam et al. proposed a systematic method to construct a quantum circuit corresponding to a QUBO or HUBO problem \cite{gilliam2021grover} with integer coefficients.
In this circuit, addition and subtraction correspond to phase advance and delay, respectively, and the objective function value is expressed by the two's complement.\footnote{It is stated in \cite{gilliam2021grover} that Gilliam's construction method is similar to the quantum Fourier transform adder \cite{draper2000addition}. 
Given the other modern quantum adders such as \cite{gidney2018halving}, there exist other methods to construct the Grover oracle.}
With the aid of this two's complement representation, the state in which the objective function value becomes negative can be identified by only a single Z gate.
After the pioneering work by Gilliam et al. \cite{gilliam2021grover}, an extension to real coefficients \cite{norimoto2023quantum} and methods of reducing constant overhead \cite{giuffrida2022engineering, yukiyoshi2022quantum,zhu2022realizable} have been considered.

A quantum circuit of GAS requires approximately $n+m$ qubits,\footnote{To be precise, some ancillae are required but not dominant.} where $n$ is the number of binary variables and $m$ is the number of qubits required for representing the objective function value.
That is, the number of qubits can be reduced by reducing the number of binary variables and restricting the value range of the objective function.
In addition, the number of quantum gates is mainly determined by the number of qubits and the number of terms in the objective function without factorization, as exemplified in \cite{gilliam2021grover} and its open-source implementation available in Qiskit.

In conventional studies \cite{sawaya2020resourceefficient, TABI2020QUANTUM, fuchs2021efficient, salehi2022unconstrained, GLOS2022SPACEEFFICIENT, domino2022quadratic}, innovative formulations for combinatorial optimization problems such as the traveling salesman problem (TSP) and the graph coloring problem (GCP) have been considered to reduce the number of binary variables.
Specifically, Tabi et al. mapped GCP to a higher-order optimization problem of a Hamiltonian using binary encoding and reduced the number of qubits required for QAOA \cite{TABI2020QUANTUM}.
In addition, Glos et al. formulated TSP as a HUBO problem and reduced the number of qubits \cite{GLOS2022SPACEEFFICIENT}.
Then, the authors combined both QUBO and HUBO formulations, demonstrating the reduction in the number of gates compared with the HUBO only case \cite{GLOS2022SPACEEFFICIENT}.
Following the pioneering studies \cite{TABI2020QUANTUM,GLOS2022SPACEEFFICIENT}, in the channel assignment problem (CAP), Sano et al. devised a HUBO formulation with different binary encoding method and reduced the numbers of qubits and gates required for GAS \cite{SANO2023QUBIT}.
Both methods \cite{GLOS2022SPACEEFFICIENT,SANO2023QUBIT} of reducing the numbers of qubits and gates are promising when compared with a simple HUBO formulation; however, both involve an increase in the number of gates compared with the original QUBO formulation.
Ideally, both the numbers of qubits and gates should be reduced in terms of implementation cost, feasibility, and circuit runtime, which is a common issue that should be addressed.

Against this background, we propose a general method of reducing the numbers of qubits and gates required for GAS.
The major contributions of this paper are summarized as follows.
\begin{enumerate}
    \item We propose a strategy termed HUBO with polynomial factorization (HUBO-PF) that reduces the number of gates by allocating Gray-coded binary vectors to indices and factorizing terms in the objective function.
    The factorized terms are mapped to quantum gates, and X gates are as successive as possible, resulting in a significant decrease in the total number of gates.
    \item In addition, we propose a strategy termed HUBO with order reduction (HUBO-OR) that halves the maximum order of the objective function, at the cost of slightly increasing the number of qubits.
    \item Numerical analysis demonstrates the reduction in the numbers of qubits and gates compared with the original QUBO formulation, which is first achieved with our strategies. This reduction accelerates the convergence performance of GAS.
\end{enumerate}

The fundamental limitation of GAS is that it assumes a future realization of FTQC. In general, Grover's algorithm is sensitive to noise and it cannot provide a quantum advantage on a current quantum computer \cite{reitzner2019grover,stoudenmire2023grover}. Even under quantum-favorable assumptions, it is known that quantum-accelerated combinatorial optimization takes much longer execution time on a current small quantum computer than on a classical computer \cite{sanders2020compilation}.

The remainder of this paper is organized as follows. In Section~\ref{sec:qubit}, we review the conventional HUBO formulations for GCP and TSP as examples. In Section~\ref{sec:gate}, we propose HUBO-OR and HUBO-PF strategies.
In Section~\ref{sec:eval}, we provide theoretical and numerical evaluations for the proposed strategies.
Finally, in Section~\ref{sec:conc}, we conclude this paper.

\section{Conventional QUBO and HUBO Formulations \cite{LUCAS2014ISING,TABI2020QUANTUM,GLOS2022SPACEEFFICIENT,SANO2023QUBIT}}\label{sec:qubit}
Lucas \cite{LUCAS2014ISING} provided Ising model formulations for many NP-complete and NP-hard problems, including QUBO problems with one-hot encodings, such as TSP and GCP.
In this section, we review QUBO and HUBO formulations \cite{LUCAS2014ISING,TABI2020QUANTUM,GLOS2022SPACEEFFICIENT,SANO2023QUBIT} for TSP and GCP as representative examples. Note that some formulations are different from the original ones introduced in \cite{TABI2020QUANTUM,GLOS2022SPACEEFFICIENT,SANO2023QUBIT}, but are essentially equivalent in a broader sense.

%In the conventional studies \cite{TABI2020QUANTUM,GLOS2022SPACEEFFICIENT,SANO2023QUBIT}, the QUBO problems are reformulated as HUBO problems to reduce the number of qubits required.
%Although the formulation method of \cite{SANO2023QUBIT} is only considered for CAP, it is readily applicable to general combinatorial optimization problems.}

\subsection{QUBO to HUBO Conversion with Reduced Qubits}
A QUBO problem that relies on one-hot encoding can be reformulated as a HUBO problem with a reduced number of qubits.
Here, one-hot encoding is an encoding method that represents an activated index as a one-hot vector.
For instance, an activated index of $1$ is represented as $[1~0~0~0]$, whereas $3$ is represented as $[0~0~1~0]$. 
These vectors can alternatively be represented as binary vectors, such as $1 \rightarrow [0 ~ 0]$ and $3 \rightarrow [1 ~ 0]$. This encoding method using binary vectors is called binary encoding.
The binary encoding reduces the number of binary variables by one logarithmic order \cite{TABI2020QUANTUM,GLOS2022SPACEEFFICIENT,SANO2023QUBIT}.
In \cite{TABI2020QUANTUM,GLOS2022SPACEEFFICIENT}, binary vectors are assigned to indices in ascending order, whereas binary vectors are assigned to indices in descending order in \cite{SANO2023QUBIT}.
Later, a HUBO formulation with ascending assignment is termed \textit{HUBO-ASC}, whereas a HUBO formulation with descending assignment is termed \textit{HUBO-DSC}.

If an optimization problem has multiple indices, a target index that should be binary encoded can be identified by the logarithm of the original search space size $S$. Specifically, the number of binary variables that is sufficient to represent the whole search space is given by $\lceil \log_2 S \rceil$ \cite{GLOS2022SPACEEFFICIENT}. If this number includes a cardinality represented in the logarithm, the corresponding index can be represented in binary.

Note that, in the QUBO to HUBO conversion considered in this paper, a QUBO problem formulated without one-hot encoding, such as the set packing problem \cite{LUCAS2014ISING}, cannot be formulated as a HUBO problem.

\subsection{Example 1: TSP}
TSP is a problem of finding a route that minimizes the total travel cost when a salesman passes through all cities once and returns to the city from which the salesman started.

\paragraph*{QUBO Formulation \cite{LUCAS2014ISING, HASEGAWA2021OPTIMIZATION, GLOS2022SPACEEFFICIENT}}
Let $N$ be the number of cities.
A typical QUBO formulation with one-hot encoding is expressed as \cite{LUCAS2014ISING, HASEGAWA2021OPTIMIZATION, GLOS2022SPACEEFFICIENT}
\begin{align}
\min_{x} \quad & E_{\text{TSP}}^{\text{QUBO}}(x) = \sum_{u=1}^{N-1} \sum_{v=u+1}^{N} W_{uv} \sum_{i=1}^{N} x_{ui}x_{vi+1} \nonumber \\
& + \lambda_1 \sum_{v=1}^{N}\left(1 - \sum_{i=1}^{N} x_{vi} \right)^2 + \lambda_2 \sum_{i=1}^{N}\left(1 - \sum_{v=1}^{N} x_{vi}\right)^2 \nonumber \\
\textrm{s.t.} \quad & x_{vi} \in \mathbb{B},
\label{eq:QUBO-TSP}
\end{align}
where each of $u$ and $v$ denotes the index of the city, $i$ denotes the order of cities to be visited, and $W_{uv}$ denotes the distance between the cities.
Binary variables are defined as
\begin{align}
x_{vi} &= 
  \begin{cases}
    1\hphantom & (\text{visit $v$th city in $i$th order})\\
    0 & (\text{otherwise})
  \end{cases}
  \label{eq:TSPQUBOxvi}
\end{align}
for $1 \leq v \leq N$ and $1 \leq i \leq N$.
That is, this QUBO formulation requires $N^2$ binary variables.
In \eqref{eq:QUBO-TSP}, $\lambda_1$ and $\lambda_2$ are penalty coefficients for constraints. Specifically, $\lambda_1$ imposes a constraint that the salesman can only visit each city once and $\lambda_2$ imposes a constraint that the salesman cannot visit more than one city at the same time.

As clarified in \cite{GLOS2022SPACEEFFICIENT}, this QUBO formulation is not efficient in terms of the number of qubits required.
Since TSP has $N!$ solutions, it is sufficient to encode all possible solutions by $\lceil \log (N!) \rceil = N \log N - N \log e + \Theta(\log N)$ binary variables \cite{GLOS2022SPACEEFFICIENT}, which is smaller than $N^2$ in the QUBO formulation.

\paragraph*{HUBO Formulation \cite{GLOS2022SPACEEFFICIENT,SANO2023QUBIT}}
%In the HUBO formulation, indices of a QUBO problem are represented as binary numbers. 
The QUBO formulation of TSP has two indices: $N$ cities and its order of visits.
Since both depend on the number of cities, $N$, either is acceptable.
In this paper, the index of the order of visits is binary encoded. 

The index $i$ of the order of visits is represented as a binary vector $[b_{i1}~b_{i2}~\cdots~b_{iB}]$, where the vector length is $B = \lceil \log_{2} N \rceil$.
That is, the HUBO formulation of TSP requires $N \lceil \log_{2} N \rceil$ binary variables.
In HUBO with ascending assignment, termed HUBO-ASC, the $B$-bit sequence is defined by
\begin{align}
[b_{i1}~b_{i2}~\cdots~b_{iB}] = [i - 1]_2, \label{eq:i-1}
\end{align}
where $[ \cdot ]_2$ denotes the decimal to binary conversion.
Note that this HUBO-ASC formulation is essentially identical with that given in \cite{GLOS2022SPACEEFFICIENT}.
By contrast, in HUBO with descending assignment, termed HUBO-DSC, the $B$-bit sequence is defined by
\begin{align}
[b_{i1}~b_{i2}~\cdots~b_{iB}] = [N - i + 1]_2.
\end{align}
Note that this HUBO-DSC formulation is an analogy of \cite{SANO2023QUBIT}.
In both HUBO-ASC and HUBO-DSC cases, we have new binary variables $x_{vr}$ for $1 \leq v \leq N$ and $1 \leq r \leq B$, and a binary state where the $v$th city is visited in $i$th order is indicated by
\begin{align}
\delta_{vi}^{(B)}(x) = \prod_{r=1}^{B}
\underbrace{(1 - b_{ir} + (2b_{ir} - 1) x_{vr})}_{x_{vr}~\textrm{if}~b_{ir}=1,~ (1-x_{vr})~\textrm{if}~b_{ir}=0},
\label{eq:deltavi}
\end{align}
which is equivalent to the binary indicator variable \eqref{eq:TSPQUBOxvi} of the QUBO case.

\begin{table}[tb]
	\centering
	\caption{Example of binary encodings in HUBO-ASC/DSC. \label{table:ch}}
	\footnotesize
	\begin{tabular}{lll}
	    \hline
		 & HUBO-ASC & HUBO-DSC\\
		$i$ & $[b_{i1} ~ b_{i2}],~~\delta_{vi}^{(2)}(x)$ & $[b_{i1} ~ b_{i2}],~~\delta_{vi}^{(2)}(x)$\\
		\hline
		$1$ & $[0 ~ 0],~~(1-x_{v1})(1-x_{v2})$ & $[1 ~ 1],~~ \,\,\,\,\,\,\,\,\,\,\,\,\,x_{v1}~\,\,\,\,\,\,\,\,\,\,\,\,\,x_{v2}$ \\
		$2$ & $[0 ~ 1],~~(1-x_{v1}) \,\,\,\,\,\,\,\,\,\,\,\,\,x_{v2}$ & $[1 ~ 0],~~ \,\,\,\,\,\,\,\,\,\,\,\,\,x_{v1}~(1-x_{v2})$ \\
		$3$ & $[1 ~ 0],~~\,\,\,\,\,\,\,\,\,\,\,\,\,x_{v1}~(1-x_{v2})$ & $[0 ~ 1],~~ (1-x_{v1})\,\,\,\,\,\,\,\,\,\,\,\,\,x_{v2}$ \\
        $4$ & $[1 ~ 1],~~\,\,\,\,\,\,\,\,\,\,\,\,\,x_{v1}~\,\,\,\,\,\,\,\,\,\,\,\,\,x_{v2}$ & $[0 ~ 0],~~ (1-x_{v1})(1-x_{v2})$ \\
		\hline
	\end{tabular}
\end{table}
As an example, for TSP with $N=4$ cities, Table~\ref{table:ch} shows the relationship between binary vectors $[b_{i1} ~ b_{i2}]$ and the state $\delta_{vi}^{(2)}(x)$ for the index $i$.
In the HUBO-ASC case, $[x_{v1} ~ x_{v2}] = [0 ~ 0]$ indicates a state where the $v=1$st city is visited in the $i=1$st order, and $\delta_{11}^{(2)}(x)$ becomes $1$. Similarly, in the HUBO-DSC case, $[x_{v1} ~ x_{v2}] = [1 ~ 1]$ indicates the same state.
As given, the total number of terms in $\delta_{vi}^{(2)}(x)$ remains the same for HUBO-ASC and -DSC since we have $N=4$.
If $N$ is not a power of 2, HUBO-DSC reduces the total number of terms and simplify the corresponding quantum circuit. That is, in Table~\ref{table:ch}, if $N=3$, the term $(1 - x_{v1})(1 - x_{v2})$ for $i=4$ is not included in the objective function, and the total number of terms is reduced.

The objective function that represents the traveling cost is given by
\begin{align}
E_{\text{TSP1}}^{\text{HUBO}}(x) = \sum_{u=1}^{N-1} \sum_{v=u+1}^{N} W_{uv} \sum_{i=1}^{N} 
\delta_{ui}^{(B)}(x) \delta_{vi+1}^{(B)}(x). \label{eq:E1}
\end{align}
To impose the constraint that the salesman cannot visit more than one city at the same time, we add
\begin{align}
E_{\text{TSP2}}^{\text{HUBO}}(x) = \sum_{i=1}^{N} \left(1 - \sum_{v=1}^{N} \delta_{vi}^{(B)}(x) \right)^2. \label{eq:E2}
\end{align}
Additionally, if $N<2^{B}$, we add a penalty function to impose the constraint that the order of visits must be less than or equal to $N$, i.e.,
\begin{align}
E_{\text{TSP3}}^{\text{HUBO}}(x) = \sum_{v=1}^{N} \sum_{i=N+1}^{2^{B}} \delta_{vi}^{(B)}(x). \label{eq:E3}
\end{align}
Overall, from \eqref{eq:E1}, \eqref{eq:E2}, and \eqref{eq:E3}, our HUBO-ASC or -DSC formulation of TSP is given by
\begin{align}
\begin{aligned}
\min_{x} \quad & E_{\text{TSP}}^{\text{HUBO}}(x) = E_{\text{TSP1}}^{\text{HUBO}}(x) + \lambda_1' E_{\text{TSP2}}^{\text{HUBO}}(x) + \lambda_2' E_{\text{TSP3}}^{\text{HUBO}}(x) \\
\textrm{s.t.} \quad & x_{vr} \in \mathbb{B},
\end{aligned}
\end{align}
where $\lambda_1'$ and $\lambda_2'$ denote the penalty coefficients for imposing the constraints.

\subsection{Example 2: GCP}
Given an undirected graph $G=(\cV, \cE)$ and a set of colors $\cI$, GCP is a problem of coloring vertices $\cV$ or edges $\cE$.
Later, the cardinalities of $\cV$, $\cE$, and $\cI$ are denoted by $V=|\cV|$, $E=|\cE|$, and $I=|\cI|$, respectively. In this paper, we consider a vertex coloring problem in which adjacent vertices are painted with different colors.

\paragraph*{QUBO Formulation \cite{LUCAS2014ISING,TABI2020QUANTUM}}
A typical QUBO formulation for GCP is expressed as \cite{LUCAS2014ISING}
\begin{align}
\begin{aligned}
\min_{x} \quad & E_{\text{GCP}}^{\text{QUBO}}(x) = \sum_{(u,v) \in \cE} \sum_{i=1}^{I} x_{ui}x_{vi} \\
& \quad \quad \quad \quad \quad + \lambda \sum_{v=1}^{V}\left(1 - \sum_{i=1}^{I} x_{vi} \right)^2 \\
\textrm{s.t.} \quad & x_{vi} \in \mathbb{B}, \label{eq:gcp_qubo}
\end{aligned}
\end{align} 
where each of $u$ and $v$ denotes vertex indices, $i$ denotes color indices, and $\lambda$ denotes the penalty coefficient for imposing a constraint: each vertex can only be painted with one color.
Binary variables are defined as
\begin{align}
x_{vi} &= 
  \begin{cases}
    1\hphantom & \text{($v$th vertex is painted with $i$th color)}\\
    0 & \text{(otherwise)}.
  \end{cases}
  \label{eq:GCPQUBOxvi}
\end{align}

\paragraph*{HUBO Formulation \cite{TABI2020QUANTUM,SANO2023QUBIT}}
The search space size of the original GCP is given by
\begin{align}
    S_{\mathrm{GCP}} = 
    {I}^{V},
    \label{eq:searchspace_GCP}
\end{align}
and $\lceil \log_2(S_{\mathrm{GCP}}) \rceil = \lceil V \log_2 I \rceil$ binary variables are sufficient to represent the whole set of solutions.
Then, the color indices should be represented as binary vectors in the HUBO formulation. The number of bits required to represent the color index is
\begin{align}
B' = \lceil \log_{2} I \rceil. \label{eq:nb'}
\end{align}
In HUBO-ASC, $B'$ bits are assigned to color indices in ascending order, which is essentially identical with the HUBO formulation proposed in \cite{TABI2020QUANTUM}.
By contrast, in HUBO-DSC, $B'$ bits are assigned to color indices in descending order, which is an analogy of \cite{SANO2023QUBIT}.
In both cases, we have new binary variables $x_{vr}$ for $1 \leq v \leq V$ and $1 \leq r \leq B'$, and
a binary state where the $v$th vertex is painted with the $i$th color is indicated by $\delta_{vi}^{(B')}(x)$ of \eqref{eq:deltavi}, which is equivalent to the binary indicator variable \eqref{eq:GCPQUBOxvi} of the QUBO case. The relationship between $b_{ir}$ and $x_{vr}$ in $\delta_{vi}^{(B')}(x)$ is the same as that given in Table~\ref{table:ch}.
Overall, our HUBO-ASC or DSC formulation of GCP is given by
\begin{align}
\begin{aligned}
\min_{x} \quad & E_{\text{GCP}}^{\text{HUBO}}(x) = \sum_{(u,v) \in \cE} \sum_{i=1}^{I} \delta_{ui}^{(B')}(x)\delta_{vi}^{(B')}(x) \\
& ~ ~ ~ ~ ~ ~ ~ ~ ~ ~ ~ ~ ~ ~ ~  + \lambda' \sum_{v=1}^{V} \sum_{i=I+1}^{2^{B'}} \delta_{vi}^{(B')}(x) \\
\textrm{s.t.} \quad & x_{vr} \in \mathbb{B}, \label{eq:gcp_hubo}
\end{aligned}
\end{align} 
where $\lambda'$ denotes the penalty coefficient for imposing a constraint: the number of colors should be less than or equal to $I$.

\section{Proposed Gate Count Reduction Strategies}\label{sec:gate}
The previous study \cite{SANO2023QUBIT}, HUBO-DSC, reduced the number of gates by the descending assignment of binary vectors as compared to the ascending assignment, HUBO-ASC, used in \cite{TABI2020QUANTUM,GLOS2022SPACEEFFICIENT}. However, even with HUBO-DSC, the number of gates remains higher than that in the original QUBO formulation. To address this issue, in this section, we propose two novel strategies that reduce the number of quantum gates required for GAS.

\subsection{Higher-Order Formulation with Polynomial Factorization: HUBO-PF Strategy}
In the HUBO formulation with binary encoding reviewed in Section~\ref{sec:qubit}, 
an arbitrary index $1 \leq i \leq I$ is mapped to a binary vector $[b_{i1} ~ \cdots b_{i \lceil \log_2 I \rceil}]$, and each binary number corresponds to a term $(1-x)$ or $x$.
Then, the resultant objective function has many terms including $(1-x)$ and $x$.
Here, $(1-x)=0$ holds if $x=1$, and $(1-x)=1$ holds if $x=0$.
Using these relationships, HUBO-PF reduces the number of quantum gates in GAS as much as possible.

\begin{figure}[tb]
	\centering
	\subfigure[HUBO-ASC/DSC.]{
		\includegraphics[clip, scale=0.5]{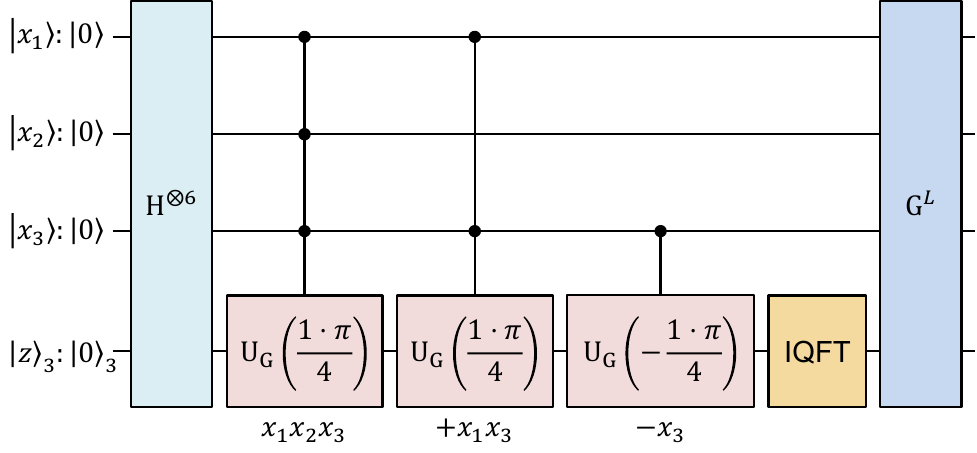}
	}
	\subfigure[HUBO-PF.]{
		\includegraphics[clip, scale=0.5]{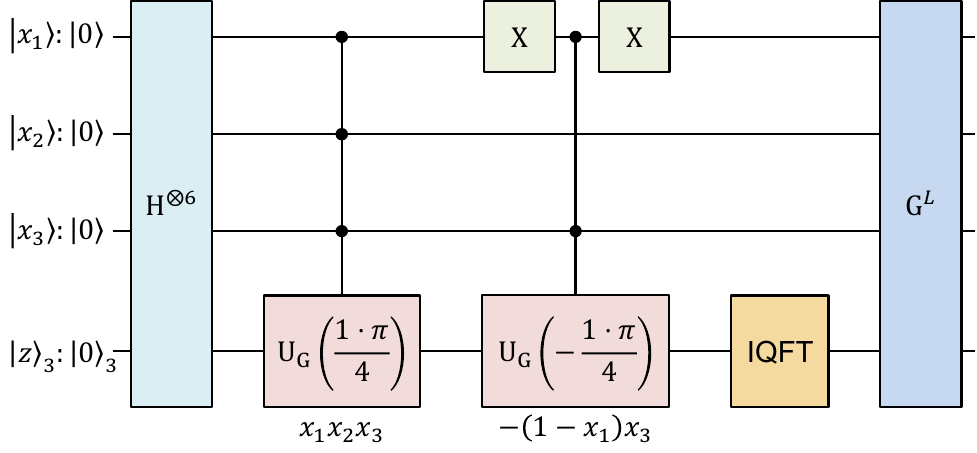}
	}
	\caption{Quantum circuits for computing $E(x)=x_{1} x_{2}x_{3} - (1-x_{1}) x_{3}$. \label{fig:circuit}}
\end{figure}
HUBO-PF exploits the gate construction method proposed by Gilliam et al. \cite{gilliam2021grover}.
Gilliam's method supports a QUBO or HUBO objective function and allows one to construct a quantum circuit in a systematic manner.
For example, if we have a HUBO-ASC or DSC function $E(x)=x_{1} x_{2}x_{3}+x_{1} x_{3}-x_{3}$, the corresponding circuit is systematically constructed as given in Fig.~\ref{fig:circuit}(a), where we have $n=3$ qubits for binary variables and $m=3$ qubits for encoding the values of the objective function.
In the beginning of Fig.~\ref{fig:circuit}(a), an equal superposition is created by the Hadamard gate $\mathrm{H}^{\otimes 6}$.
An arbitrary coefficient in the HUBO function, denoted by $a$, is represented as $\theta=2\pi a/2^m$, and the corresponding unitary operator is given by \cite{gilliam2021grover}
\begin{align}
    \mathrm{U}_{\mathrm{G}}(\theta) = \mathrm{R}(2^{m-1} \theta)
    \otimes \mathrm{R}(2^{m-2} \theta) \otimes \cdots \otimes \mathrm{R}(2^{0} \theta)
    \label{Ug-theta}
\end{align}
and the phase gate \cite{gilliam2021grover}
\begin{align}
    \mathrm{R}(\theta) = \begin{bmatrix}
        1 & 0 \\ 0 & e^{j\theta}
    \end{bmatrix}.\label{eq:Rtheta}
\end{align}
Then, the first third-order term $1 \cdot x_{1} \cdot x_{2} \cdot x_{3}$ corresponds to the controlled-$\mathrm{U}_G(1 \cdot \pi / 4)$ in Fig.~\ref{fig:circuit}(a), which is controlled by $\ket{x_{1}}$, $\ket{x_{2}},$ and $\ket{x_{3}}$.
The definition of the Grover operator $\mathrm{G}$ is detailed in \cite{gilliam2021grover}, and it is applied $L$ times to amplify the states of interest.
Note that an open-source implementation of GAS is available in Qiskit.

The idea of HUBO-PF is to map factorized terms directly to quantum gates.
Specifically, the original Qiskit implementation of GAS \cite{gilliam2021grover} supports an objective function without factorization and constructs a quantum circuit including gates corresponding to all the terms.
Here, there is no need to expand the objective function sequentially.
That is, since the objective functions such as $E_{\text{TSP}}^{\text{HUBO}}(x)$ and $E_{\text{GCP}}^{\text{HUBO}}(x)$ are already factorized, the factorized terms can be directly mapped to quantum gates.

We describe HUBO-PF in detail using a concrete example.
In Fig.~\ref{fig:circuit}(a), we had a HUBO-ASC/DSC function
\begin{align}
    E(x)=x_{1} x_{2}x_{3}+x_{1} x_{3}-x_{3},
\end{align}
but it could be factorized into
\begin{align}
    E(x)=x_{1} x_{2} x_{3} - (1-x_{1}) x_{3}.
\end{align}
Since the X gate is equivalent to a bit flipping, the multiplication of $(1-x_1)$ is equivalent to a multiplication of $x_1$ sandwiched between two X gates.
Then, in the factorized case, the term $- (1-x_{1}) x_{3}$ can be mapped to a controlled unitary gate sandwiched between two X gates as in Fig.~\ref{fig:circuit}(b).
As given, Fig.~\ref{fig:circuit}(a) has three $\mathrm{U}_G(\theta)$ gates, which are reduced to two in Fig.~\ref{fig:circuit}(b), although both circuits are equivalent.

\begin{figure}[tb]
	\centering
    \includegraphics[clip, scale=0.5]{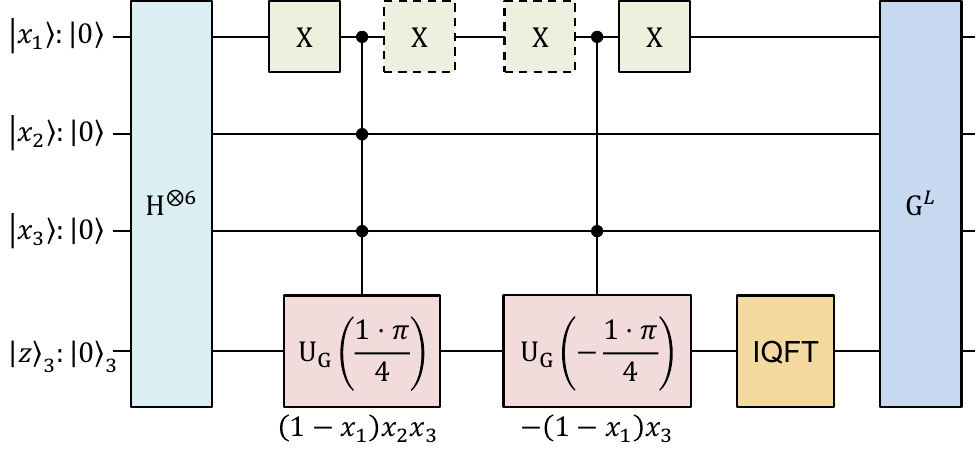}
	\caption{Quantum circuit for computing $(1-x_{1}) x_{2}x_{3} - (1-x_{1}) x_{3}$, where X gates are cancelled. \label{fig:x-circ}}
\end{figure}
In addition, successive X gates can be canceled each other.
For example, if we have another factorized function $(1-x_{1}) x_{2}x_{3} - (1-x_{1}) x_{3}$, the corresponding quantum circuit can be constructed as in Fig.~\ref{fig:x-circ}.
As given, we have two $(1-x_{1})$ terms, and originally, four X gates are required in total.
Here, the $(1-x_{1})$ terms are mapped to a circuit in the same manner as in Fig.~\ref{fig:circuit}(b), and X gates become successive.
The X gates with dotted lines in Fig.~\ref{fig:x-circ} cancel each other and can therefore be eliminated.
Other X gates that were not canceled are single qubit operations; potential parallelism with other quantum gates can be expected.

\begin{table}[tb]
	\centering
	\caption{Example of binary encoding in HUBO-PF. \label{table:ch2}}
	\small
	\begin{tabular}{llll}
	    \hline
		Index & HUBO-PF \\
        $i$ & $[b_{i1} ~ b_{i2} ~ b_{i3}]$, & $\delta_{vi}^{(3)}(x)$ \\
		\hline
		$1$ & $[1 ~ 1 ~ 1]$, & $\,\,\,\,\,\,\,\,\,\,\,\,\,x_{v1}~ \,\,\,\,\,\,\,\,\,\,\,\,\,x_{v2}~ \,\,\,\,\,\,\,\,\,\,\,\,\,x_{v3}$\\
		$2$ & $[1 ~ 0 ~ 1]$, & $\,\,\,\,\,\,\,\,\,\,\,\,\,x_{v1}~ (1-x_{v2}) \,\,\,\,\,\,\,\,\,\,\,\,\,x_{v3}$\\
		$3$ & $[1 ~ 0 ~ 0]$, & $\,\,\,\,\,\,\,\,\,\,\,\,\,x_{v1}~ (1-x_{v2})(1-x_{v3})$\\
		$4$ & $[0 ~ 0 ~ 0]$, & $(1-x_{v1})    (1-x_{v2})(1-x_{v3})$ \\
		$5$ & $[0 ~ 0 ~ 1]$, & $(1-x_{v1})    (1-x_{v2}) \,\,\,\,\,\,\,\,\,\,\,\,\,x_{v3}$\\
		$6$ & $[0 ~ 1 ~ 1]$, & $(1-x_{v1})    \,\,\,\,\,\,\,\,\,\,\,\,\,x_{v2}~ \,\,\,\,\,\,\,\,\,\,\,\,\,x_{v3}$\\
        $7$ & $[0 ~ 1 ~ 0]$, & $(1-x_{v1})    \,\,\,\,\,\,\,\,\,\,\,\,\,x_{v2}~ (1-x_{v3})$\\
		$8$ & $[1 ~ 1 ~ 0]$, & $\,\,\,\,\,\,\,\,\,\,\,\,\,x_{v1}~ \,\,\,\,\,\,\,\,\,\,\,\,\,x_{v2}~ (1-x_{v3})$\\
		\hline
	\end{tabular}
\end{table}
One fundamental question arises in this situation. How can we make X gates as successive as possible?
A solution by optimization is undesirable owing to its large overhead in solving a combinatorial optimization problem.
To make X gates as successive as possible, we can devise a binary allocation so that the bits change one by one with respect to the index.
Such an allocation method is commonly known as the Gray code in digital communications, which mitigates bit errors because the Hamming distance between adjacent bits is kept one.

In HUBO-ASC or DSC, binary vectors are assigned to indices in ascending or descending order.
In HUBO-PF, the first index is assigned to a binary vector with all ones, and subsequent binary vectors are the same as the Gray code.
In this way, $(1-x)$ terms are kept as contiguous as possible.
Table~\ref{table:ch2} shows an example of binary encoding of 3 bits in HUBO-PF that can reduce the number of X gates.
A binary vector $[1 ~ \cdots ~ 1]$ is assigned to the first index, and the following binary vectors are changed one by one, which is the same as the Gray code.
That is, with respect to the index $i$, binary numbers have many successive zeros.
In a certain objective function, a summation is calculated in ascending order with respect to the index. A quantum circuit is constructed in that order, and X gates will naturally be consecutive.
This HUBO-PF assignment minimizes the number of X gates in a heuristic manner without imposing additional optimization cost.
Note that HUBO-PF is also applicable to the QUBO case if the objective function contains many $(1-x)$ terms.
Unlike GCP and TSP, if an objective function has an irregular indexing, the HUBO-PF assignment cannot reduce the number of X gates. But, in that case, the Gray code optimized with respect to the irregular indexing would also reduces the number of X gates.

\subsection{Higher-Order Formulation with Order Reduction: HUBO-OR Strategy}
In practice, a quantum computer has limited connectivity between qubits, and this limitation imposes additional circuit latency \cite{shafaei2014qubit}.
Our HUBO-PF strategy is optimal in terms of the number of qubits and gates, although it induces high-order terms in the objective function, resulting in many controlled gates and interactions between qubits.
To circumvent this issue, we propose a strategy referred to as HUBO-OR, which serves as an intermediate strategy situated between HUBO-PF and the conventional QUBO, thereby striking a balance between the two.
HUBO-OR halves the maximum order of the objective function with the sacrifice of a small increase in the number of qubits.

The idea of HUBO-OR is to map limited binary vectors to indices.
Specifically, binary vectors that have an even number of ones, including a binary vector with all zeros, are assigned to the target indices in order to halve the maximum order.
In HUBO-OR, the first index is assigned to a binary vector with all zeros. Subsequent binary vectors, sorted in ascending order, have an even number of ones.
This strategy is particularly beneficial for HUBO problems where interference occur between the same indices, such as overlapping colors in GCP.

\begin{table}[tb]
	\centering
	\caption{Example of binary coding in HUBO-OR. \label{table:ch3}}
	\small
	\begin{tabular}{lll}
	    \hline
		Index & HUBO-OR & \\
        $i$ & $[b_{i1} ~ b_{i2} ~ b_{i3}]$, & $\delta_{vi}^{(3)}(x)$ \\
		\hline
		$1$      & $[0 ~ 0 ~ 0]$, & $(1-x_{v1})    (1-x_{v2})(1-x_{v3})$ \\
		Not used & $[0 ~ 0 ~ 1]$, & $(1-x_{v1})    (1-x_{v2}) \,\,\,\,\,\,\,\,\,\,\,\,\,x_{v3}$\\
		Not used & $[0 ~ 1 ~ 0]$, & $(1-x_{v1})    \,\,\,\,\,\,\,\,\,\,\,\,\,x_{v2}~ (1-x_{v3})$\\
		$2$      & $[0 ~ 1 ~ 1]$, & $(1-x_{v1})    \,\,\,\,\,\,\,\,\,\,\,\,\,x_{v2}~ \,\,\,\,\,\,\,\,\,\,\,\,\,x_{v3}$\\
		Not used & $[1 ~ 0 ~ 0]$, & $\,\,\,\,\,\,\,\,\,\,\,\,\,x_{v1}~ (1-x_{v2})(1-x_{v3})$\\
		$3$      & $[1 ~ 0 ~ 1]$, & $\,\,\,\,\,\,\,\,\,\,\,\,\,x_{v1}~ (1-x_{v2}) \,\,\,\,\,\,\,\,\,\,\,\,\,x_{v3}$\\
        $4$      & $[1 ~ 1 ~ 0]$, & $\,\,\,\,\,\,\,\,\,\,\,\,\,x_{v1}~ \,\,\,\,\,\,\,\,\,\,\,\,\,x_{v2}~ (1-x_{v3})$\\
        Not used & $[1 ~ 1 ~ 1]$, & $\,\,\,\,\,\,\,\,\,\,\,\,\,x_{v1}~ \,\,\,\,\,\,\,\,\,\,\,\,\,x_{v2}~ \,\,\,\,\,\,\,\,\,\,\,\,\,x_{v3}$\\
		\hline
	\end{tabular}
\end{table}
Let us check an example of HUBO-OR formulation. In the GCP case, we have $I$ colors, and the number of bits required to represent the color index is
\begin{align}
B'' = \lceil \log_{2} I \rceil + 1, \label{eq:nb''}
\end{align}
rather than $\lceil \log_{2} I \rceil$, since we only use the limited binary vectors, which have an even number of ones.
The number of bits $B''$ is large enough that the total of available binary vectors is calculated as
\begin{align}
\sum_{r=0}^{\lfloor B'' / 2 \rfloor} {B'' \choose 2r} = 2^{\lceil \log_{2} I \rceil} \geq I
\end{align}
for $I \geq 1$.
For example, if we have $I = 4$ colors, $B'' = \lceil \log_{2} 4 \rceil + 1 = 3$ bits are sufficient to represent the color index.
Table~\ref{table:ch3} shows a binary encoding of 3 bits in HUBO-OR.
In this formulation, the first index is assigned to the all-zero binary vector, and the following indices are assigned to binary vectors that have an even number of ones.
For new binary variables $x_{vr}$ for $1 \leq v \leq V$ and $1 \leq r \leq B''$, a binary state where the $v$th vertex is painted with the $i$th color is indicated by $\delta_{vi}^{(B'')}(x)$ of \eqref{eq:deltavi}, where the relationship between the index $i$, $b_{ir}$, and $x_{vr}$ in $\delta_{vi}^{(B'')}(x)$ is given in Table~\ref{table:ch3}.

Similar to the first term of the HUBO-ASC/DSC formulation \eqref{eq:gcp_hubo}, the objective function that represents the interference between vertices is given by
\begin{align}
E_{\text{GCP1}}^{\text{HUBO-OR}}(x) = \sum_{(u,v) \in \cE} \prod_{r=1}^{B''} (1 - x_{ur} - x_{vr}),
\end{align}
which is always positive since each of the binary vectors contains even number of ones.
Another method to make it positive is to compute the square $(1 - x_{ur} - x_{vr})^2$, but this computation is not desirable in terms of order reduction.
To impose the constraint that no index is assigned to a binary vector having an odd number of $1$, we add
\begin{align}
E_{\text{GCP2}}^{\text{HUBO-OR}}(x) = \sum_{v=1}^{V} \sum_{i=1}^{I} \left( H_i \bmod{2} \right) \delta_{vi}^{(B'')}(x),
\end{align}
where $H_i$ denotes the Hamming weight of an index $i$, i.e.,
\begin{align}
    H_i = \sum_{r=1}^{B''} b_{ir}.
\end{align}
Additionally, we add a penalty function to impose the constraint that the number of colors to be painted is less than or equal to $I$ if $I<2^{B''-1} = 2^{\lceil \log_{2} I \rceil}$, i.e.,
\begin{align}
E_{\text{GCP3}}^{\text{HUBO-OR}}(x) = \sum_{v=1}^{V} \sum_{i=I+1}^{2^{B''}} \delta_{vi}^{(B'')}(x).
\end{align}
Overall, our HUBO-OR formulation of GCP is given by
\begin{align}
\begin{aligned}
&\min_{x}  \quad  E_{\text{GCP}}^{\text{HUBO-OR}}(x) = E_{\text{GCP1}}^{\text{HUBO-OR}}(x)  \\
&~~~~~~~~~~~~~~~~~~~~~ + \lambda_1'' E_{\text{GCP2}}^{\text{HUBO-OR}}(x)\\
&~~~~~~~~~~~~~~~~~~~~~ + \lambda_2'' E_{\text{GCP3}}^{\text{HUBO-OR}}(x) \\
&\textrm{s.t.} \quad  x_{vr} \in \mathbb{B}, 
\end{aligned}\label{eq:gcp_half}
\end{align}
where $\lambda_1''$ and $\lambda_2''$ denote penalty coefficients. 

According to the HUBO-ASC/DSC formulation of \eqref{eq:gcp_hubo}, the maximum order is calculated as $2B' = 2\lceil \log_{2} I \rceil$.
By contrast, according to the HUBO-OR formulation of \eqref{eq:gcp_half}, the maximum order is calculated as $B'' = \lceil \log_{2} I \rceil + 1$, which is almost halved.

Note that HUBO-OR and HUBO-PF can be combined in some cases.
In the case of GCP, HUBO-PF is applicable to the constraint terms $E_{\text{GCP2}}^{\text{HUBO-OR}}(x)$ and $E_{\text{GCP3}}^{\text{HUBO-OR}}(x)$ because of the sequential references to index $i$ in summation. But, in $E_{\text{GCP2}}^{\text{HUBO-OR}}(x)$, we have limited cases where $H_i \bmod{2} = 1$ is satisfied successively for increasing $i$, and in $E_{\text{GCP3}}^{\text{HUBO-OR}}(x)$, we have limited terms from $i=I+1$ to $2^{B''}$. Then, it is also considered that the positive effect of gate count reduction could be limited. The combination of HUBO-PF and HUBO-OR strategies may require additional optimization to significantly reduce the number of gates.

\section{Algebraic Analysis and Numerical Evaluation}\label{sec:eval}
In this section, we analyze the conventional QUBO, HUBO-ASC \cite{TABI2020QUANTUM,GLOS2022SPACEEFFICIENT}, and HUBO-DSC \cite{SANO2023QUBIT} as well as the proposed HUBO-PF and HUBO-OR in terms of the numbers of qubits and quantum gates, where GCP is considered as an example.\footnote{Our proposed strategies are applicable to other QUBO problems relying on one-hot encoding.}
Note that we omit the discussion on the depth of quantum circuit, because it heavily depends on heuristic algorithms used in the transpilation process and the assumed connectivity of a quantum computer, leading to different conclusions depending on specific assumptions.

The search space size of the original GCP is given by
\begin{align}
    S_{\mathrm{GCP}} = 
    {I}^{V},
\end{align}
indicating that the optimal number of binary variables should be
\begin{align}
    \underline{n} = \log_2(S_{\mathrm{GCP}}) = V \log_2 I.
\end{align}
Then, if the number of binary variables becomes much larger than $\underline{n}$, that formulation may not be efficient in terms of the number of qubits.

\subsection{Analysis on Number of Qubits}
Firstly, we analyze the number of qubits required for GAS.
GAS requires approximately $n+m$ qubits \cite{gilliam2021grover}, where $n$ is the number of binary variables and $m$ is the smallest integer satisfying \cite{yukiyoshi2022quantum}
\begin{align}
-2^{m-1} \leq \min[E(x)] \leq \max[E(x)] < 2^{m-1}, \label{eq:m2}
\end{align}
owing to the two's complement representation of the 
objective function.\footnote{In practice, additional qubits are required owing to the gate decomposition and the connectivity of a quantum device.}

\paragraph*{QUBO}
From \eqref{eq:gcp_qubo}, the conventional QUBO formulation for GCP requires
\begin{align}
    n=VI
    \label{eq:n}
\end{align}
binary variables.
When considering an equal superposition of $2^n$ states, the size of search space is
\begin{align}
    C = 2^n = 2^{VI},
\end{align}
which is larger than $S_{\mathrm{GCP}}$ for most typical parameters.
From \eqref{eq:gcp_qubo}, the objective function value is always positive and its maximum is 
\begin{align}
\text{max}(E_{\text{GCP}}^{\text{QUBO}}(x)) =EI + \lambda V(I - 1)^2. \label{eq:max_qubo}
\end{align}
Then, the number of qubits required to encode $E_{\text{GCP}}^{\text{QUBO}}(x)$ is
\begin{align}
m = \lceil \log_{2} (\text{max}(E_{\text{GCP}}^{\text{QUBO}}(x)))\rceil,
\label{eq:m}
\end{align}
since $E_{\text{GCP}}^{\text{QUBO}}(x)$ itself is always positive.
The total number of required qubits is calculated as\footnote{We use $\Omega(\cdot)$ for asymptotic lower bounds.}
\begin{align}
n + m =  VI +  \left\lceil \log_{2} \left( EI + \lambda V(I - 1)^2 \right) \right\rceil = \Omega(VI).\label{EQ:NM}
\end{align}

\paragraph*{HUBO-ASC/DSC}
In the HUBO-ASC/DSC case, we have
\begin{align}
    n'=V \lceil \log_{2} I \rceil
    \label{eq:n'}
\end{align}
binary variables and the size of search space is
\begin{align}
    C' = 2^{n'} = 2^{V \lceil \log_{2} I \rceil}
    = \left(2^{\log_2 I + O(1)}\right)^V = \Omega(I^V),
    \label{eq:C'}
\end{align}
which is almost equal to the original size of search space $S_{\mathrm{GCP}}$.
From \eqref{eq:gcp_hubo}, the maximum value of the objective function is
\begin{align}
\text{max}(E_{\text{GCP}}^{\text{HUBO}}(x)) = E. \label{eq:max_hubo}
\end{align}
Then, since the objective function value is always positive, the number of qubits required to encode $E_{\text{GCP}}^{\text{HUBO}}(x)$ is
\begin{align}
m' = \lceil \log_{2} (\text{max}(E_{\text{GCP}}^{\text{HUBO}}(x))\rceil. \label{eq:m'}
\end{align}
The total number of required qubits is calculated as
\begin{align}
n' + m' =  V \lceil \log_{2} I \rceil + \left\lceil \log_{2} E \right \rceil = \Omega(V \log_2 I). \label{EQ:NM'}
\end{align}

\paragraph*{HUBO-PF}
Note that HUBO-PF requires the same number of qubits as the HUBO-ASC/DSC case since it relies on the equivalent objective function of \eqref{eq:gcp_hubo}, where terms are not expanded. The size of search space is also the same as that in \eqref{eq:C'}.

\paragraph*{HUBO-OR}
In the HUBO-OR case, we have
\begin{align}
    n''=V (\lceil \log_{2} I \rceil + 1)
    \label{eq:n''}
\end{align}
binary variables, and the size of search space is
\begin{align}
    2^{n''} = 2^{V (\lceil \log_{2} I \rceil + 1)} = \Omega(I^V),
\end{align}
which is the same as that in the HUBO-ASC/DSC case.
From \eqref{eq:gcp_half}, the maximum value of the objective function is
\begin{align}
\text{max}(E_{\text{GCP}}^{\text{HUBO-OR}}(x)) = E, \label{eq:max_half}
\end{align}
and the number of qubits required to encode $E_{\text{GCP}}^{\text{HUBO-OR}}(x)$ is
\begin{align}
m'' = \lceil \log_{2} (\text{max}(E_{\text{GCP}}^{\text{HUBO-OR}}(x))\rceil  = m'. \label{eq:m''}
\end{align}
The total number of required qubits is calculated as
\begin{align}
n'' + m'' &=  V (\lceil \log_{2} I \rceil + 1) + \left\lceil \log_{2} E \right \rceil = \Omega(V (\log_2 I + 1)). \label{EQ:NM''}
\end{align}

\begin{figure}[tb]
	\centering
    \includegraphics[clip, scale=0.68]{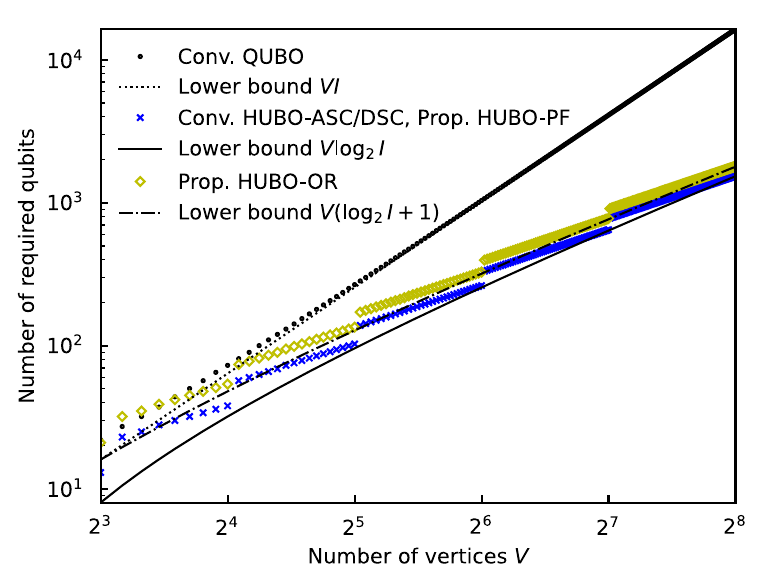}
	\caption{Numbers of qubits required by the conventional and proposed formulation methods, where markers indicate actual numbers and lines indicate their asymptotic lower bounds given in \eqref{EQ:NM}, \eqref{EQ:NM'}, and \eqref{EQ:NM''}.\label{fig:qubit}}
\end{figure}
Fig.~\ref{fig:qubit} shows the number of qubits required by the conventional and proposed formulation methods, where we set $I=V / 4$ and $E=3V$.
In addition, all the penalty coefficients are set to 1.
As shown in Fig.~\ref{fig:qubit}, the derived asymptotic lower bounds, given in \eqref{EQ:NM}, \eqref{EQ:NM'}, and \eqref{EQ:NM''}, were almost identical to the actual number of qubits calculated if the number of vertices $V$ was sufficiently large.
HUBO-ASC/DSC/OR required significantly fewer qubits than the conventional QUBO, indicating that the proposed HUBO formulation can reduce the size of search space.

\subsection{Analysis of Quantum Gate Count}\label{subsec:anagate}
Secondly, we analyze the number of quantum gates, which is an important evaluation metric that determines the feasibility of a quantum circuit.
In this paper, we construct concrete quantum circuits using Gilliam's construction method \cite{gilliam2021grover} and count the specific numbers of quantum gates.
In the following, we analyze the number of gates corresponding to the state preparation operator $\mathrm{A}_y$ in GAS \cite[Eq.~(4)]{gilliam2021grover}, which has a dominant impact on the constructed circuits of GAS.

\paragraph*{QUBO}
In the conventional QUBO formulation, the number of H gates is
\begin{align}
    G_{\mathrm{H}} = n+m = \Omega(VI).
\end{align}
The numbers of R and controlled-R (CR) gates can be calculated using \eqref{eq:m}.
Specifically, the number of CR gates corresponds to the number of first-order terms in the objective function and is calculated as
\begin{align}
    G_{\textrm{CR}} = VI m = \Omega(VI \log_2 (\lambda VI^2)).\label{eq:QUBO1CR}
\end{align}
%where we have
%\begin{align}
%\alpha = \left\lceil \log_{2} \left(EI + \lambda V(I - 1)^2 \right)  \right\rceil +1.
%\end{align}
Similarly, the number of controlled-controlled-R (CCR or $\textrm{C}^2\textrm{R}$) gates is calculated as
\begin{align}
G_{\textrm{CCR}} &= \left(EI + V \binom{I}{2} \right) \cdot m =\frac{2EI + VI(I-1)}{2} \cdot m \nonumber \\
&= \Omega(VI^2 \log_2 (\lambda VI^2)).
\label{eq:QUBO2CR}
\end{align}

\paragraph*{HUBO-ASC}
In the HUBO-ASC formulation, the number of H gates is
\begin{align}
    G_{\textrm{H}}' = n' + m' = \Omega(V \log_2 I). \label{eq:HUBOH}
\end{align}
Similar to the QUBO case, the numbers of R and multiple-qubit $\textrm{C}^k\textrm{R}$ gates can be calculated from \eqref{eq:m'}, where $1 \leq k \leq B'$ denotes the number of control qubits.
We calculate the number of $k$th order terms, which is given by
\begin{align}
G_{\textrm{C}^k\textrm{R}}' &= \left\{ E \binom{2B'}{k} - \binom{B'}{k} \sum_{v=1}^{V} \mathrm{deg}(v)-1 \right\} \cdot m' \nonumber \\
&= \Omega\left(\frac{(2 \log_2 I)^k}{k!} E \log_2 E \right),\label{eq:HUBOkCR1}
\end{align}
where $\mathrm{deg}(v)$ denotes the degree of the vertex $v$.
%, and we have
%\begin{align}
%    \alpha' = \left\lceil \log_{2} E \right\rceil +1.
%\end{align}
Additionally, the number of $\textrm{C}^k\textrm{R}$ gates for $B' < k \leq 2B'$ is calculated as
\begin{align}
G_{\textrm{C}^k\textrm{R}}' = E \binom{2B'}{k} \cdot m' = \Omega\left(\frac{(2 \log_2 I)^k}{k!} E \log_2 E \right). \label{eq:HUBOkCR2}
\end{align}

\paragraph*{HUBO-DSC}
In the HUBO-DSC formulation, it is not possible to derive the numbers of gates in using closed-form expressions.
The basic trend remains the same as in the HUBO-ASC case, but the number of gates can be reduced, which depends on the problem parameters.

\paragraph*{HUBO-PF}
In the HUBO-PF formulation, a quantum circuit is constructed using the factorized objective function as in \eqref{eq:gcp_hubo}.
The analysis of gate count poses a certain degree of complexity.
The number of H gates is the same as \eqref{eq:HUBOH}.
According to \eqref{eq:m'}, the number of CR gates is calculated as
\begin{align}
G_{\textrm{C}^{B'}\textrm{R}}' &= V(2^{B'} - I) \cdot m' \nonumber \\
&= V (2^{\log_2 I + O(1)} - I) (\log_2 E + O(1)) \nonumber \\
&= \Omega\left(V \log_2 E \right)
\end{align}
and
\begin{align}
G_{\textrm{C}^{2B'}\textrm{R}}' = EI \cdot m' = \Omega\left(EI \log_2 E \right),
\end{align}
since the objective function consists of only $B'$-order and $2B'$-order terms.
In addition, HUBO-PF introduces the use of additional X gates and the gate count is calculated as
\begin{align}
2^{B'} V B' = \Omega\left(V I \log_2 I \right),\label{eq:x_conv}
\end{align}
which can be further reduced to
\begin{align}
G_{\mathrm{X}}' = 2^{B'} V = \Omega\left(V I \right)\label{eq:x_prop}
\end{align}
by canceling successive X gates.
That is, as exemplified in Table~\ref{table:ch2}, the X gates become maximally successive by using the simple Gray code, thereby reducing the number of gates significantly.

\paragraph*{HUBO-OR}
In the HUBO-OR formulation, according to \eqref{EQ:NM''}, the number of H gates is
\begin{align}
    G_{\textrm{H}}'' = n'' + m'' = \Omega(V \log_2 I).
\end{align}
From \eqref{eq:m''}, for a general $1 \leq k \leq B''$, the number of $\textrm{C}^k\textrm{R}$ gates is calculated as
\begin{align}
G_{\textrm{C}^k\textrm{R}}'' &= \left\{2^k E \binom{B''}{B''-k} - \binom{B''}{k} \sum_{v=1}^{V} (E_{v}-1)\right\} \cdot m' \nonumber \\
&= \Omega\left(\frac{(2 \log_2 I)^k}{k!} E \log_2 E\right).
\end{align}

\subsection{Numerical Evaluation of Quantum Gate Count}
To evaluate the conventional and proposed formulation methods, we assume that $I=V / 4$ and $E=3V$ hold upon increasing the number of vertices $V$.
Moreover, we set $\mathrm{deg}(v) = 6$ and all the penalty coefficients to 1.
Note that in the following figures, in Figs.~\ref{fig:terms}--\ref{fig:cdf}, markers were added to make each line easier to distinguish, which had no special intentions.

\begin{figure}[tb]
	\centering
    \includegraphics[clip, scale=0.68]{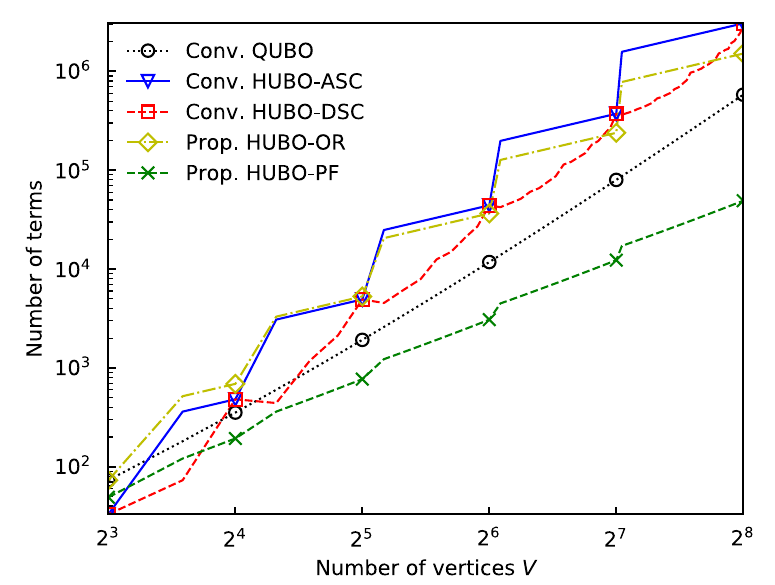}
	\caption{Actual numbers of terms required by the conventional and proposed formulation methods.\label{fig:terms}}
\end{figure}
First, Fig.~\ref{fig:terms} shows the actual number of terms in the objective function of each formulation method.
Here, we considered the QUBO, HUBO-ASC, HUBO-DSC, and HUBO-OR formulations, where the corresponding terms were expanded, while the terms were not expanded in the HUBO-PF case.
We calculated the actual number of terms in the HUBO-DSC case and used the derived exact count in other cases given in Section~\ref{subsec:anagate}.
Specifically, in the QUBO case, the number of terms can be calculated from \eqref{eq:QUBO1CR} and \eqref{eq:QUBO2CR} as
\begin{align}
    G_{\textrm{CR}} + G_{\textrm{CCR}} = 
    \alpha I \left( E + \frac{V (I + 1)}{2}
    \right).
\end{align}
Similarly, actual numbers of terms were calculated in other HUBO-ASC, HUBO-OR, and HUBO-PF cases.
As shown in Fig.~\ref{fig:terms}, when compared with HUBO-ASC, HUBO-DSC reduced the number of terms in most cases because it generated fewer terms in the form of $(1-x)$.
Although HUBO-ASC/DSC/OR exhibited a larger number of terms than QUBO, HUBO-PF succeeded in reducing the number of terms significantly.

Next, we count the number of T gates, which is an important metric when assuming surface-code-based FTQC.
Specifically, we count the number of T gates used in the state preparation operator $\mathrm{A}_y$ in GAS \cite[Eq.~(4)]{gilliam2021grover} and ignore the gates required for the quantum Fourier transform. The phase gate $\mathrm{R}(\theta)$ of \eqref{eq:Rtheta} may yield numerous T gates depending on the value of $\theta$. For instance, $\mathrm{R}(\pi/4) = \mathrm{T}$ and $\mathrm{R}(\pi/2) = \mathrm{T}\mathrm{T}$ hold true, but using the Solovay-Kitaev algorithm \cite{dawson2005solovaykitaev}, $\mathrm{R}(\pi/3)$ is decomposed into 84 H gates and 99 T gates, with a square error of about $1.20 \cdot 10^{-3}$. The implementation cost of $\mathrm{R}(\theta)$ is considered constant, irrespective of each formulation method, and its decomposition does not affect the relative comparison. The corresponding T-count increases with $O(m)$ but is negligible compared to the increase in $O(n)$. Then, in the following, we assume that $\mathrm{R}(\theta)$ is not decomposed and its T-count is ignored.

When assuming that $\mathrm{R}(\theta)$ is not decomposed, according to \cite{nielsen2010quantum}, the number of T gates required to decompose $\textrm{C}^n\textrm{R}$ for $n \geq 2$ into $\textrm{CR}$ is $14(n-1)$ with $n-1$ auxiliary qubits (or ancillae).
Then, a $\textrm{CR}$ gate is decomposed into CNOT and single-qubit unitary gates including $\mathrm{R}(\theta)$.
In the QUBO, HUBO-ASC/DSC (=HUBO-PF), and HUBO-OR formulations, the required numbers of ancilla qubits are
\begin{align}
&a = 1 \\
&a' = 2 \lceil \log_{2} I \rceil - 1, ~ \mathrm{and} \\
&a'' = \lceil \log_{2} I \rceil,
\end{align}
respectively.
Here, it is clear that the number of ancilla qubits is negligible compared with the total number of qubits $n + m$ for each formulation method.
The number of gates can be further reduced by using the relative-phase Toffoli (RTOF) gate, which approximates the function of the Toffoli gate. Specifically, the number of T gates required to decompose an arbitrary $n$ qubit unitary operator is $14(n-1)$ when the original Toffoli gates are used \cite{nielsen2010quantum}, whereas Maslov's approach \cite{maslov2016advantages} reduces it to $8(n-1)$ using RTOF gates.

\begin{figure}[tb]
	\centering
    \includegraphics[clip, scale=0.68]{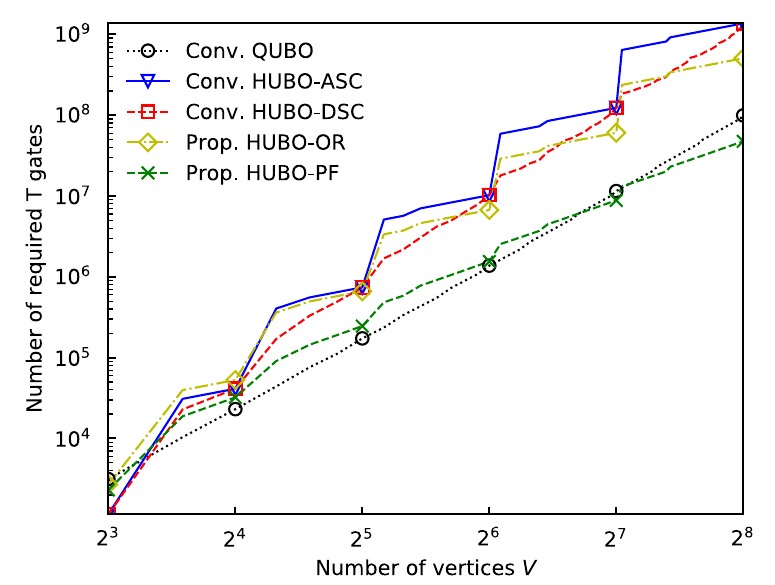}
	\caption{Estimated numbers of T gates required by the conventional and proposed formulation methods.\label{fig:tgate}}
\end{figure}
Then, Fig.~\ref{fig:tgate} shows the estimated number of T gates required by each formulation method, where we used $G_{\textrm{C}^k\textrm{R}}$, $G_{\textrm{C}^k\textrm{R}}'$, and $G_{\textrm{C}^k\textrm{R}}''$ expressions given in Section~\ref{subsec:anagate}. In the HUBO-DSC case, we counted $\textrm{C}^k\textrm{R}$ gates numerically.
As shown in Fig.~\ref{fig:tgate}, HUBO-DSC and HUBO-OR reduced the number of T gates in most cases compared with HUBO-ASC.
For a large problem size, surprisingly, HUBO-PF had fewer T gates than QUBO.
This indicates that HUBO-PF is the best formulation  strategy in terms of the numbers of qubits and T gates.

\begin{figure}[tb]
	\centering
    \includegraphics[clip, scale=0.68]{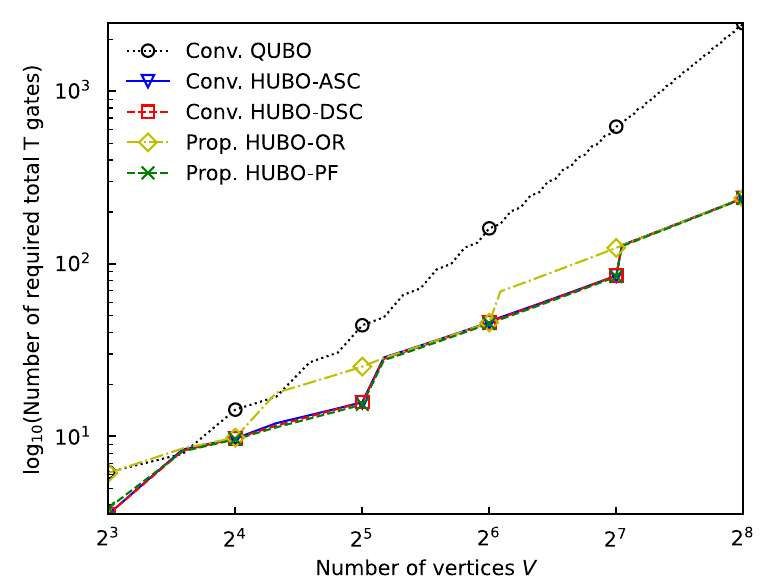}
	\caption{Estimated number of total T gates required to obtain the optimal solution.\label{fig:tgate_total}}
\end{figure}
For reference, we evaluated the estimated number of total T gates required for obtaining the optimal solution, where the estimated number of T gates shown in Fig.~\ref{fig:tgate} was used and the total number of Grover operators was assumed to be $\sqrt{2^n}$, $\sqrt{2^{n'}}$, or $\sqrt{2^{n''}}$ for each formulation.
As shown in Fig.~\ref{fig:tgate_total}, the proposed HUBO formulations reduced the total number of T gates compared with the conventional QUBO in almost all cases.

\subsection{Numerical Evaluation of Convergence Performance}
Finally, we evaluated the convergence performance of GAS using each formulation method, where we set $(V, I)= (5, 4)$.
\begin{figure}[tb]
	\centering
	\includegraphics[clip, scale=0.68]{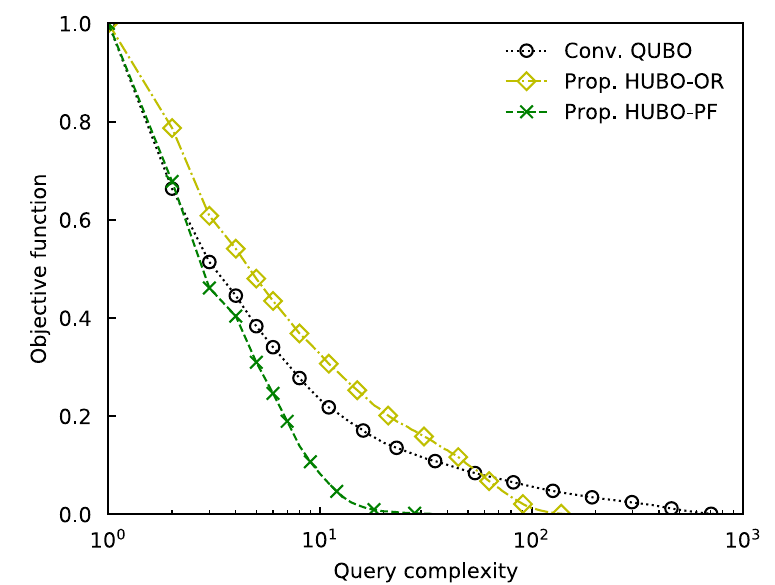}
	\caption{Transition of objective function values with respect to query complexity.\label{fig:query}} 
\end{figure}
\begin{figure}[tb]
	\centering
	\includegraphics[clip, scale=0.68]{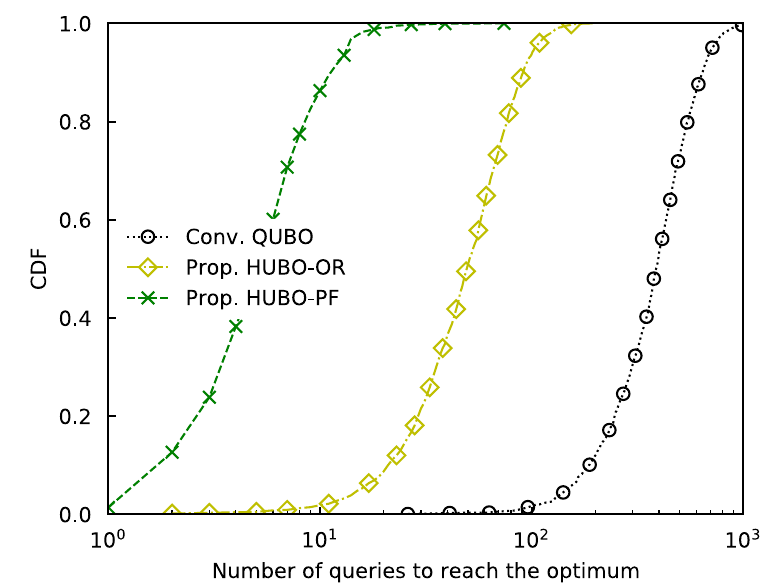}
	\caption{CDF of the query complexity required to obtain the optimal solution. \label{fig:cdf}}
\end{figure}
Fig.~\ref{fig:query} shows the transition of objective function values of QUBO, HUBO-OR, and HUBO-PF, where the number of Grover operators required to reach the optimal solution was calculated and the objective function values were normalized within $[0, 1]$.
Compared with QUBO and HUBO-OR, HUBO-PF converged to the optimal solution the fastest on average.
Note that HUBO-ASC/DSC achieves the same query complexity as the HUBO-PF case, but it is expected that HUBO-PF further reduces the circuit runtime due to its reduction in gate count.

In addition, Fig.~\ref{fig:cdf} shows the cumulative distribution function (CDF) of the query complexity required to converge to the optimal solution.
That is, Fig.~\ref{fig:cdf} is a different representation of Fig.~\ref{fig:query}.
As shown in Fig.~\ref{fig:cdf}, HUBO-PF converged the fastest with almost 100\% probability.
The sizes of search space were $2^n = 2^{VI} = 2^{20}$, $2^{n'} = 2^{V \lceil \log_2 I \rceil} = 2^{10}$, and $2^{n''} = 2^{V \lceil \log_2 I \rceil} = 2^{15}$ for QUBO, HUBO-OR, and HUBO-PF, respectively.
This reduction in the search space size leads to a significant speedup.

In Figs.~\ref{fig:query} and \ref{fig:cdf}, the query complexity of the QUBO case may be improved if we use the Dicke state \cite{bartschi2022shortdepth} instead of creating an equal superposition state by the Hadamard gates.
Nevertheless, the proposed HUBO-PF strategy still has advantages in terms of the number of qubits and gates, subsequently decreasing circuit runtime and implementation cost. In addition, the proposed HUBO-OR is an intermediate strategy between QUBO and HUBO-PF, which is suitable for a quantum computer that has limited connectivity.

\section{Conclusion}\label{sec:conc}
We reviewed HUBO-ASC/DSC formulations that effectively reduces the number of qubits required for GAS, using TSP and GCP as representative examples.
Then, we proposed the novel strategies: one that decreased the number of gates through polynomial factorization, termed HUBO-PF, and the other that halved the order of the objective function, termed HUBO-OR.
Our analysis demonstrated that the proposed strategies enhance the convergence performance of GAS by both decreasing the search space size and minimizing the number of quantum gates. Our strategies are particularly beneficial for general combinatorial optimization problems using one-hot encoding.

\footnotesize{
	\bibliographystyle{IEEEtranURLandMonthDiactivated}
	\bibliography{main}
}

\vspace{8pt}

\begin{IEEEbiographynophoto}{Yuki~Sano}
received the B.E. degree in engineering science from Iwate University, Iwate, Japan, in 2021, and the M.E. degree in engineering science from Yokohama National University, Kanagawa, Japan, in 2023. He is currently working at Nomura Research Institute based in Tokyo.
\end{IEEEbiographynophoto}

\begin{IEEEbiographynophoto}{Kosuke~Mitarai} is an Associate Professor at Osaka University, a position he has held since 2023. He completed his doctoral program in the Graduate School of Engineering Science at the same university in 2020. Since 2020, he served as an assistant professor before his current role. In 2018, he co-founded QunaSys, a quantum software startup. His research primarily focuses on developing algorithms for applying quantum computers to real-world problems.
\end{IEEEbiographynophoto}

\begin{IEEEbiographynophoto}{Naoki~Yamamoto} is a Professor at the Department of Applied Physics and Physico-Informatics, Keio University, and the chair of Keio Quantum Computing Center. He received the B.S. degree in engineering and the M.S. and Ph.D. degrees in information physics and computing from the University of Tokyo in 1999, 2001, and 2004, respectively. He was a postdoctoral fellow at the California Institute of Technology from 2004 to 2007, and at the Australian National University from 2007 to 2008. His research interests include quantum computation and control.
\end{IEEEbiographynophoto}

\begin{IEEEbiographynophoto}{Naoki Ishikawa} is an Associate Professor with the Faculty of Engineering, Yokohama National University, Kanagawa, Japan. He received the B.E., M.E., and Ph.D. degrees from the Tokyo University of Agriculture and Technology, Tokyo, Japan, in 2014, 2015, and 2017, respectively. In 2015, he was an academic visitor with the School of Electronics and Computer Science, University of Southampton, UK. From 2016 to 2017, he was a research fellow of the Japan Society for the Promotion of Science. From 2017 to 2020, he was an assistant professor in the Graduate School of Information Sciences, Hiroshima City University, Japan. He was certified as an Exemplary Reviewer of \textsc{IEEE Transactions on Communications} in 2017 and 2021. His research interests include quantum algorithms and wireless communications.
\end{IEEEbiographynophoto}

\end{document}